\documentclass[twocolumn,iop]{openjournal}
\usepackage{graphicx}
\usepackage{txfonts}
\usepackage{algorithm}
\usepackage{algpseudocode}
\usepackage{hyperref}

\begin{document}

\journalinfo{} %The Open Journal of Astrophysics}
\submitted{submitted February 2022}

\title{Coronal Mass Ejection image edge detection in Heliospheric Imager STEREO SECCHI data}
\shorttitle{Coronal Mass Ejection image edge detection}

\author{Marc D. Nichitiu}
\affiliation{The Stony Brook School, NY, USA}

\begin{abstract}
  We present an algorithm to detect the outer edges of Coronal Mass Ejection (CME) events as seen in
  differences of Heliospheric Imager STEREO SECCHI HI-1 images from either A or B spacecraft,
  as well as its implementation in Python.
\end{abstract}

\maketitle

\section{Introduction}

Studying and monitoring solar activity, and in particular,
its stronger radiation and particle emission,
is important for both science and practical purpose, for our life on and around Earth. In particular,
Coronal Mass Ejections (CME), which carry magnetized plasma,
can induce geomagnetic storms on our planet, and also endanger astronauts or space equipment. 
Increased interest in studying their origins, structure, and trajectories resulted in more and
more international space missions, among which several ones
are still operational. The STEREO pair A and B of NASA spacecraft, out of which STEREO-B was lost in 2014,
is one of such latest missions, continuously imaging the Sun and the space between Sun and Earth,
with several instruments, as described in \cite{Eyles2009SoPh..254..387E,Eyles10.1117/12.732822}
%Socker et al., 2000; Defise et al., 2003; Harrison, Davis, and Eyles, 2005;
by orbiting around the Sun at roughly the same distance as the
Earth. Early years of raw and
  processed image data are available from both, and STEREO-A is still operational.
Two visible-light cameras HI-1 and HI-2 on each spacecraft are thus
precisely aimed in order to cover the line Sun-Earth, while being shielded from the Sun itself.
Their images show various parts of the solar corona,
receiving free-electron
scattered light from the K-corona, dust scattered light from the F-corona, as well as light from background stars.
Many studies of these images start from running differences, as described by
\cite{Sheeley1997ApJ...484..472S} for coronagraph images, to enhance the faint K-corona light and remove
the slowly-varying but much stronger F-corona light, as also explained in \cite{Davies2009GeoRL..36.2102D}.
CME perturb the K-corona, and thus appear as bright regions in these images. Successive images, obtained
from space missions such as the SOHO LASCO \cite{2004JGRA..109.7105Y}, and STEREO coronagraph and HI cameras,
show their propagation and evolution. Considerable
research studying the origin of CMEs, shape, and trajectories is in progress, but
open problems remain, as described in the very comprehensive review from \cite{WebbHowardCMEObsReview2012},
in particular regarding propelling forces and interaction with solar wind: such models can be
found in \cite{XieGopalswamy2006SpWea...410002X,MichalekGopalswamy2006SpWea...410003M,Gopalswamy2007,Howard2008}. 
Given the large amount of data as well as the need for tracking, computerized pre-analysis tools
can provide important help, in particular to identify the CME boundaries and kinematics.

Computer algorithms able to identify boundaries of various objects in images have been already studied. They can use gradient techniques such as in \cite{Canny1986}, or active contour techniques:
\cite{Caselles1993,ChanVese2001,MarquezNeilaMorphActiveContours2014}.
However, as \cite{Young2008SoPh..248..457Y} explain,
it is not straightforward to use these existing techniques because of the diffuse nature of CMEs. Moreover,
they also have complex inner features, and their signal can be affected by the presence of bright stars and even the Milky Way passing in the field of view,
which adds to the noise.

Thus, several complex algorithmic methods specialized for this task have  been devised:
CACTus \cite{Robbrecht2004A&A...425.1097R},
SEEDS \cite{Olmedo2008AGUSMSP43A..02O}, and multiscale wavelet analysis
\cite{Young2008SoPh..248..457Y,Byrne2009A&A...495..325B},
mainly focused on the coronagraph images. These methods isolate
the regions of interest in specific ways, and then estimate geometric and kinematic parameters,
using underlying assumptions such as constant acceleration and elliptical shapes. 

The algorithm presented here only focuses on the image segmentation stage, identifying the
outer edge of CME regions in STEREO HI images using a staged approach composed of simple steps.
The algorithm starts by filtering and smoothing the image difference, subsequently
exploring it from the center of the bright CME region, marking its outer edges.

Through subsequent work, it can be integrated into various kinematic estimation schemes. Given
its modular nature, the algorithm itself could also be further adapted to examine specific
internal features apparent in the images.

The rest of the paper is organized as follows. In Section~\ref{Section:Algo} we describe
the algorithm outline and its subsequent processing of an image.
In Section~\ref{Section:Results} we show
its results on several other STEREO HI-1 images of known CME,
and in Section~\ref{Section:Discussion} we discuss
future lines of work.

\section{Algorithm}\label{Section:Algo}

\begin{figure}[H]
\begin{center} % Background-removed 0-to-255 normalized difference image of a 
  \includegraphics[scale=0.4]{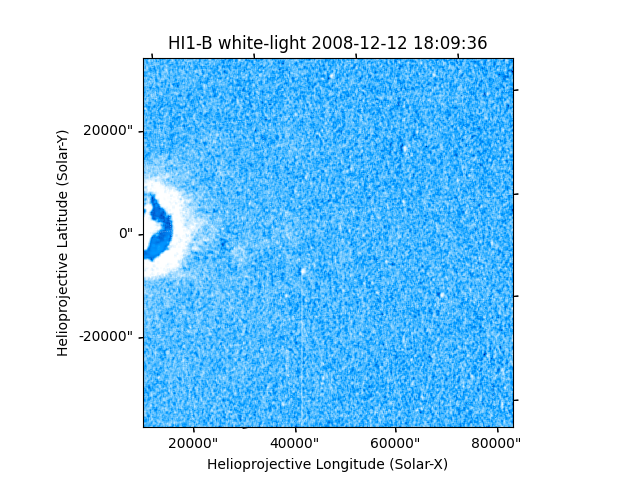}
  \caption{2008 CME event image (STEREO HI-1 Spacecraft B) used as input. The traveling CME appears as a bright region entering the field of vision on the left. The image is obtained by normalizing the difference of two consecutive background-filtered images to a 0-255 range of pixel values.}\label{Fig:PrepInput}
\end{center}
\end{figure}

In Figure~\ref{Fig:PrepInput} we see an example of what the algorithm
starts with, which is an early phase of a CME visible
in the differenced image from December 12, 2008.
The Sun is behind the right edge of the image, which has $n_c\times n_r=1024\times1024$ pixels.

\begin{figure}[H]
  \begin{center}
    \includegraphics[scale=0.138]{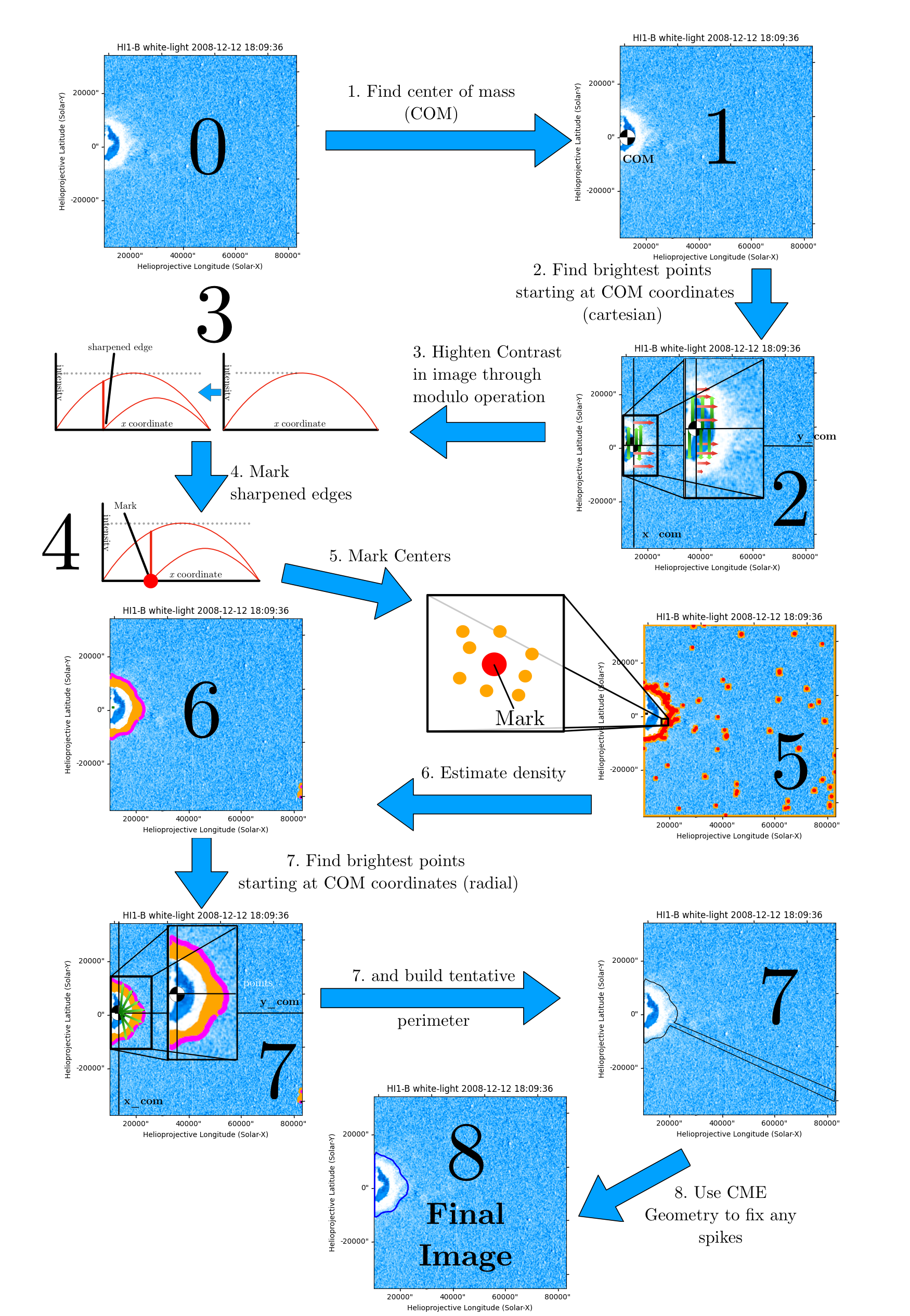}
\end{center}
\caption{Diagram illustrating the different steps of the algorithm}\label{Fig:IntermediateStages}
\end{figure}

The main steps, graphically illustrated in Figure~\ref{Fig:IntermediateStages},
are as follows, where $I[c,r]$ represents the pixel intensity
for coordinates column $c$ and row $r$ of the input image. The origin is the bottom left of the image.

\begin{enumerate}
\item\label{AlgoStep:FindCenter} Find the center of mass $(x_S,y_S)$ of the bright area and set $C=(c_S,y_S)$ where $c_S$ is close to the image edge $c_0$ towards the Sun (right for spacecraft A, so typically $c_0=1023$, and left for B, i.e. $c_0=0$).
\item\label{AlgoStep:MarkFarthestBright} Going from $c_S$ to $c_E$ (1023 for B, 0 for A),
  on each image row $r$,
  mark the farthest bright pixels seen along. Then do the same
  on each image column $c$, first from $y_S$ up, and then from $y_S$ down, with $t_1=240$ for 0-255 gray level images.
  \begin{algorithmic}[1]
\State $e_1\gets []$
\For{each image row $r$}
\State $q\gets $False
\State $v\gets c_0$
\For{each image column $c$ from $c_S$ to $c_E$}
\If{$I[c,r]>t_1$}
\State $v\gets c$
\State $q\gets $True
\EndIf
\EndFor
\If{$q$}
\State $e_1+\gets[v,r]$
\EndIf
\EndFor
\For{each image column $c$}
\For{each image row $r$ from $y_S$ to $r_E$}
\If{$I[c,r]>t_1$}
\State $v\gets r$
\State $q\gets $True
\EndIf
\EndFor
\If{$q$}
\State $e_1+\gets[c,v]$
\EndIf
\For{each image row $r$ from $y_S$ to $0$}
\If{$I[c,r]>t_1$}
\State $v\gets r$
\State $q\gets $True
\EndIf
\EndFor
\If{$q$}
\State $e_1+\gets[c,v]$
\EndIf
\EndFor
\end{algorithmic}

\item\label{AlgoStep:CreateModContrast} Smooth the image $I$ into $I_M$,
  using a sum-scale-and-modulo-256 5x5 filter to enhance contrast at the edges
\item\label{AlgoStep:MarkModContrast} Mark the contrasting pixels created in step~\ref{AlgoStep:CreateModContrast}, with $t_2=120$ for 0-255 gray level images
    \begin{algorithmic}[1]
\State $e_2\gets []$
\For{each image row $r$}
\For{each image column $c$ except the last one}
\If{$|I_M[c,r]-I_M[c+1,r]|>t_2$}
\State $e_2+\gets[c,r]$
\EndIf
\EndFor
\EndFor
\For{each image column $c$}
\For{each image row $r$ except the last one}
\If{$|I_M[c,r]-I_M[c,r+1]|>t_2$}
\State $e_2+\gets[c,r]$
\EndIf
\EndFor
\EndFor
\end{algorithmic}

\item\label{AlgoStep:IntersectContrast} Mark the centers where both kinds of previous marks are present in a $w_i\times w_i$ window, as created in step~\ref{AlgoStep:MarkFarthestBright} and step~\ref{AlgoStep:MarkModContrast}, namely from the $e_1$ and $e_2$ lists:
   \begin{algorithmic}[1]
     \State $I_1=\textrm{ones}[n_c,n_r]$
     \State $I_2=\textrm{ones}[n_c,n_r]$
\For{each element $[x,y]$ of $e_1$}
\State $I_1[x,y]\gets 2$
\EndFor
\For{each element $[x,y]$ of $e_2$}
\State $I_2[x,y]\gets 3$
\EndFor
\State $I_3\gets I_1+I_2$
\State $I_4\gets \textrm{filter}(I_3,\textrm{prod},w_i)$
\State $e_3\gets$ list of $[x,y]$ such that $I_4[x,y]$ is a multiple of $2$ and $3$, or of $5$
\end{algorithmic}

 \item\label{AlgoStep:Density} Estimate the density of the markers from step~\ref{AlgoStep:IntersectContrast} in a larger window $w_d\times w_d$ and mark those above half into a boolean mask $I_D$.
   
 \item\label{AlgoStep:CircularSweep} Circularly sweeping from the center, for each such radial half-line gather the last mark as created in step~\ref{AlgoStep:Density}, i.e.
   from $I_D$, and build the perimeter:
   \begin{algorithmic}[1]
     \State $e_4\gets[]$
     \For{each angle $\alpha$ from $0$ to $2\pi$ in steps of $\varepsilon$}
     \State $v\gets$Nothing
     \For{each pixel $[x,y]$ along the $\alpha$ ray from $[c_S,y_S]$ to any edge of the image}
     \If{$I_D[x,y]$}
     \State $v\gets[x,y]$
     \EndIf
     \EndFor
     \If{$v$}
     \State $e_4+\gets v$
     \EndIf
     \EndFor
   \end{algorithmic}
   
 \item\label{AlgoStep:RemoveOutliers} Remove spikes (outliers) created in step~\ref{AlgoStep:CircularSweep} in the list $e_4$ of pairs of pixel coordinates, reconnect and validate the final perimeter, based on CME geometry from \cite{Fisher1984CMEEvents,Fisher1984CMEGeom,Crifo1983LoopOrBubble}.

\end{enumerate}

The idea behind step~\ref{AlgoStep:CreateModContrast} is to augment existing gradual
contrasting zones by creating an abrupt transition through the modulo operation.

Given the larger density of markers around the true edge, even if the rays in step~\ref{AlgoStep:CircularSweep} are one or two pixels wide, they should not miss them when they
cross it outwards.

Outliers appear occasionally because of brighter background light points far away
from the CM, and can be removed in step~\ref{AlgoStep:RemoveOutliers}. The validation of the
perimeter, to eliminate unphysical results, is based on CME general geometry,
using bubble and loop models as described in \cite{Crifo1983LoopOrBubble}. Specifically, as
\cite{Fisher1984CMEEvents} and \cite{Fisher1984CMEEvents} point out, the width to depth ratio is
typically 3:2, so we compute and compare these diameters.

\section{Results}\label{Section:Results}

\begin{figure}
\begin{center}
\hbox{
  \includegraphics[scale=0.3]{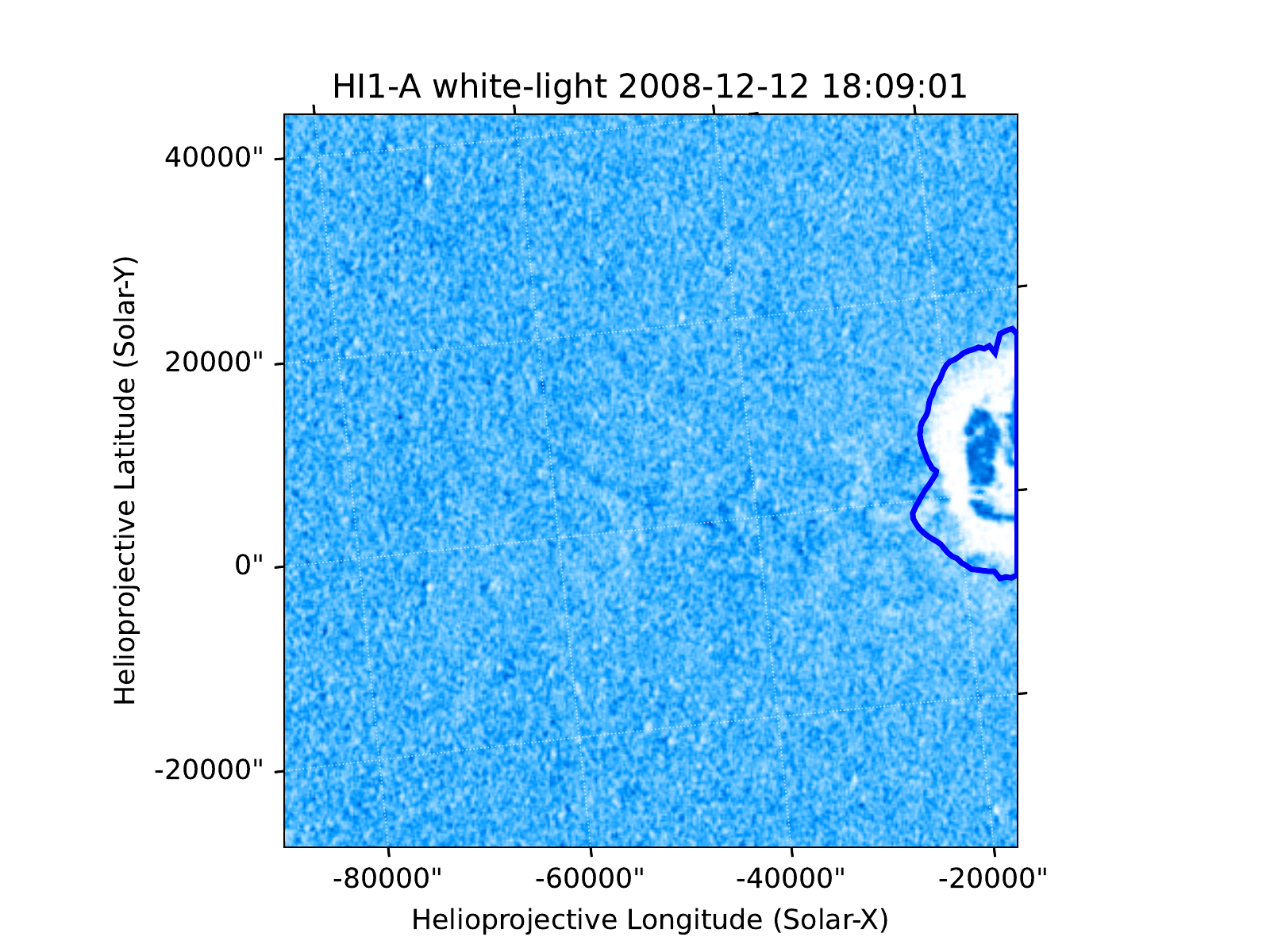}
  \kern-1.2cm
  \includegraphics[scale=0.3]{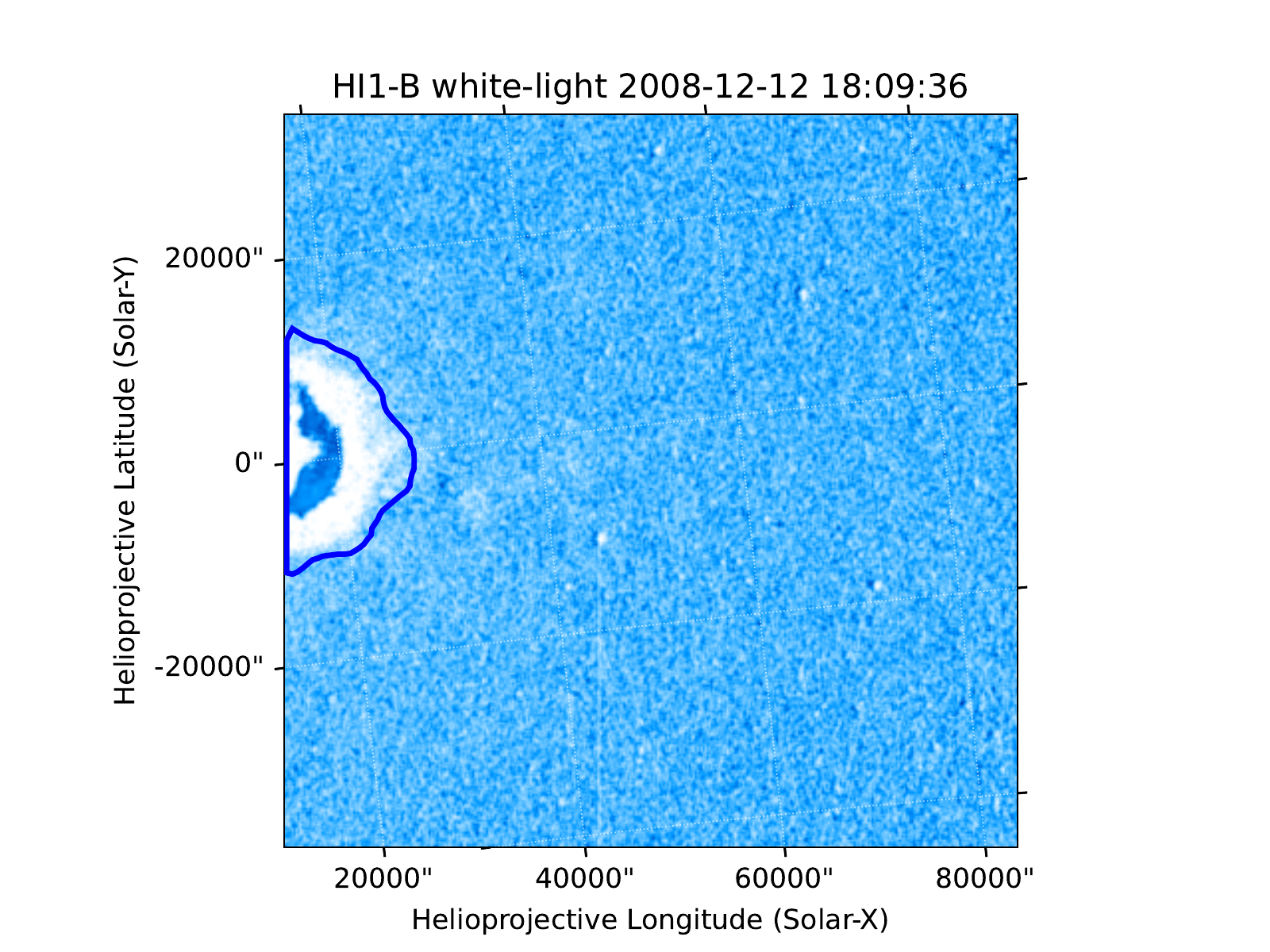}
}
\end{center}

\begin{center}
\hbox{
  \includegraphics[scale=0.3]{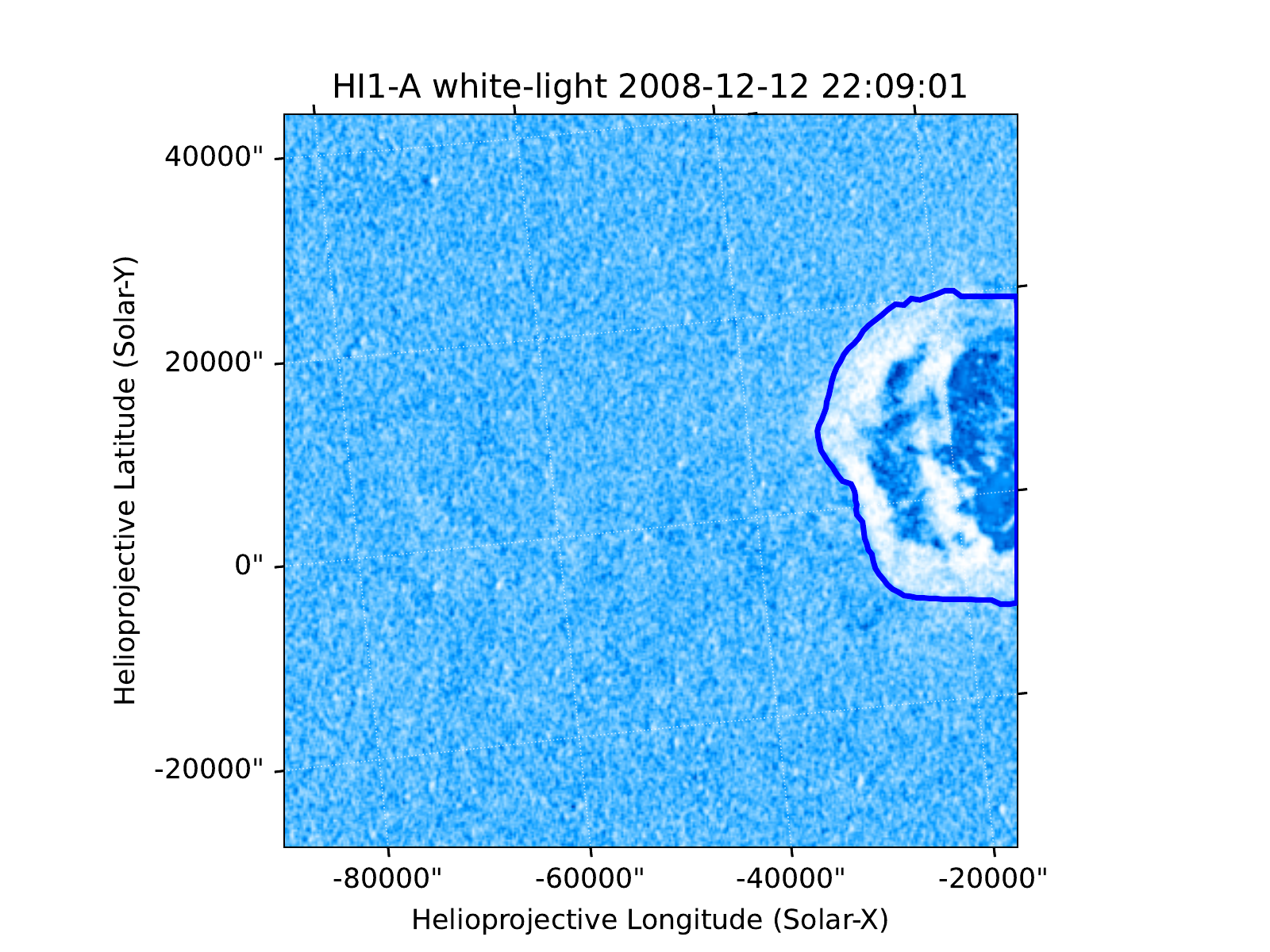}
  \kern-1.2cm
  \includegraphics[scale=0.3]{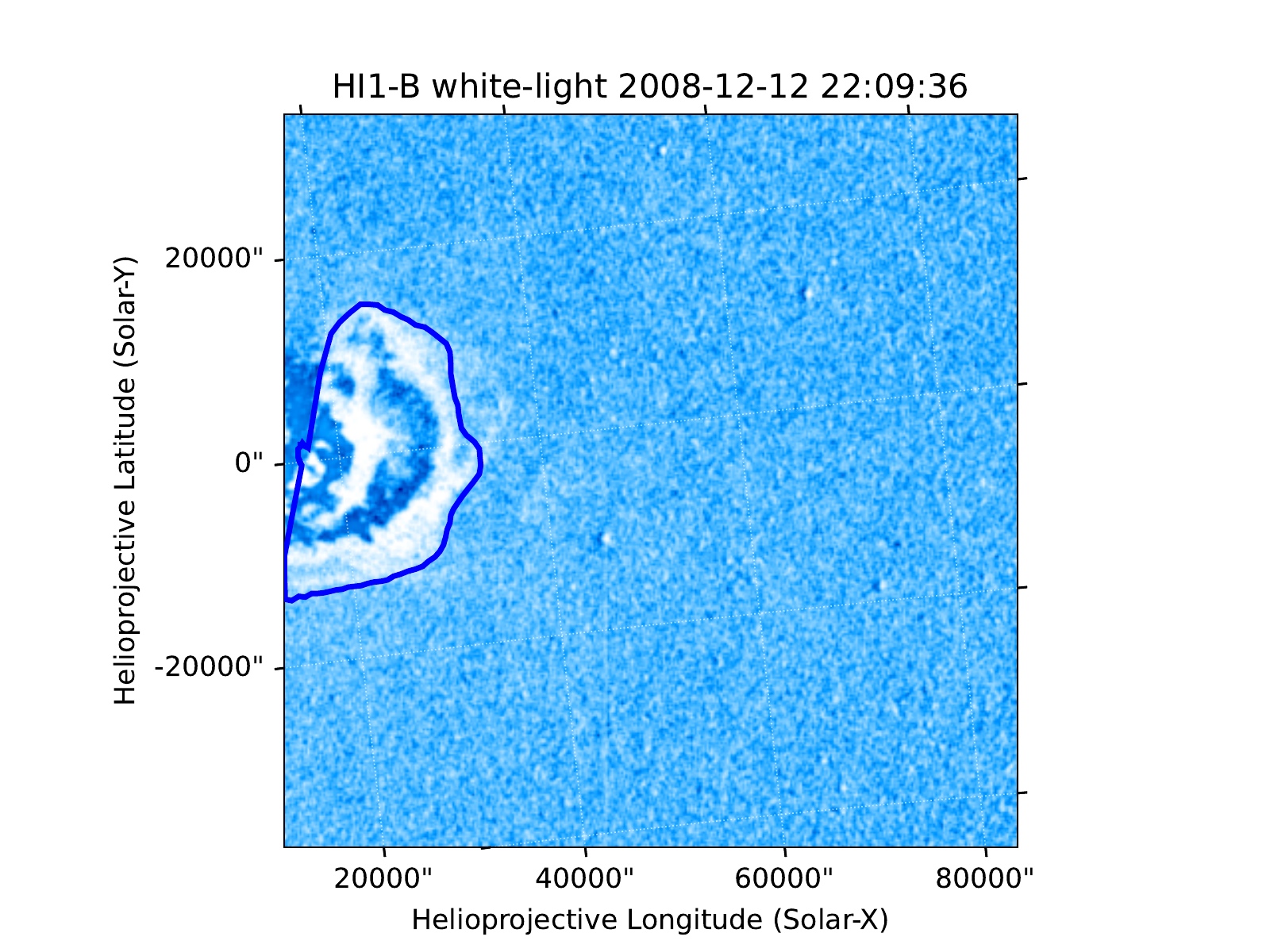}
}
\end{center}

\begin{center}
\hbox{
  \includegraphics[scale=0.3]{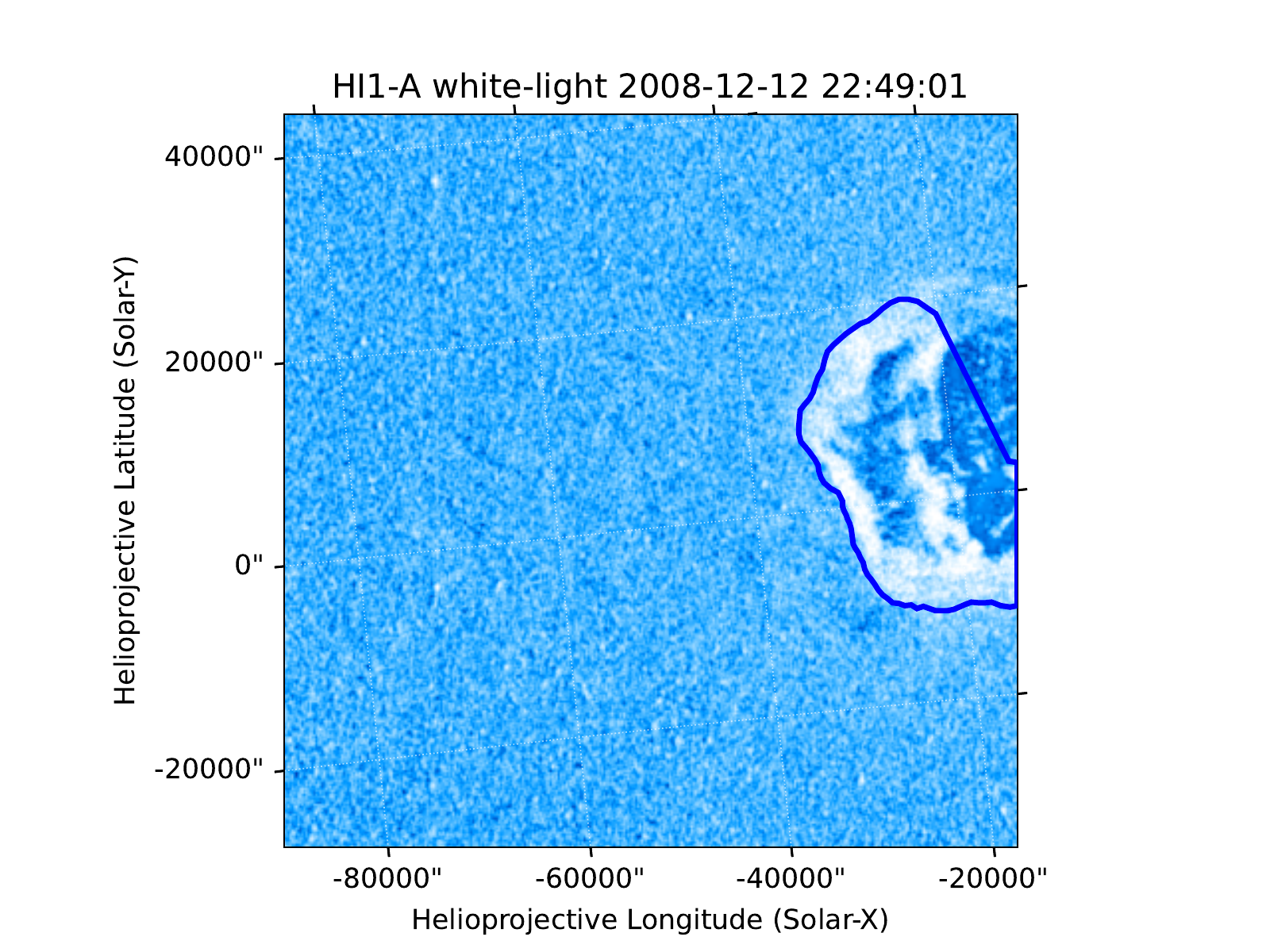}
  \kern-1.2cm
  \includegraphics[scale=0.3]{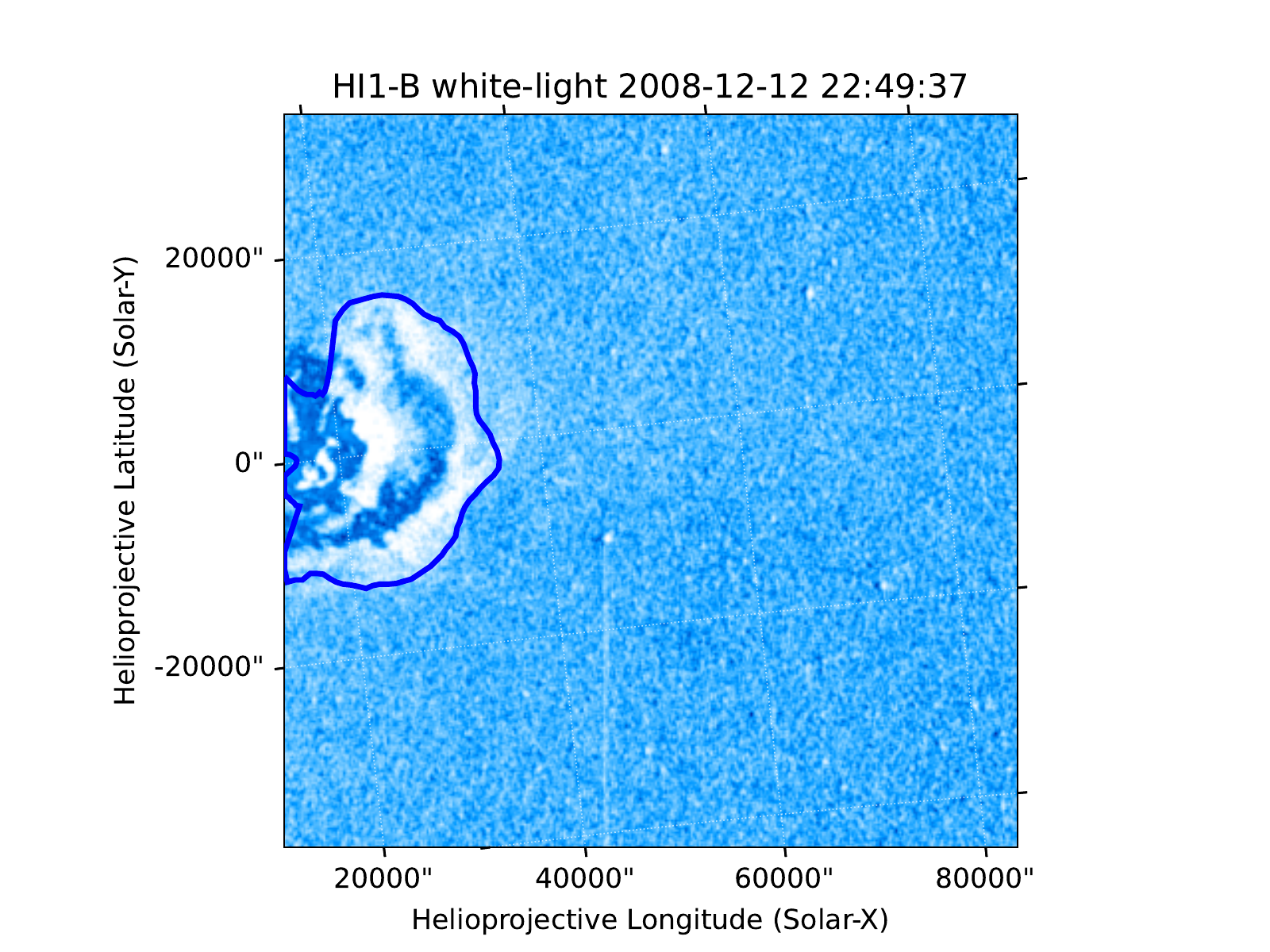}
}
\caption{A few examples of processing results: the program recognized the CME bright region boundary and marked it with dark pixels, forming a contour line. We can notice that inner darker regions do not prevent the program from finding the outermost edges.}\label{Fig:ProcExamples}
\end{center}
\end{figure}

\begin{figure}
\begin{center}
\hbox{
  \includegraphics[scale=0.2]{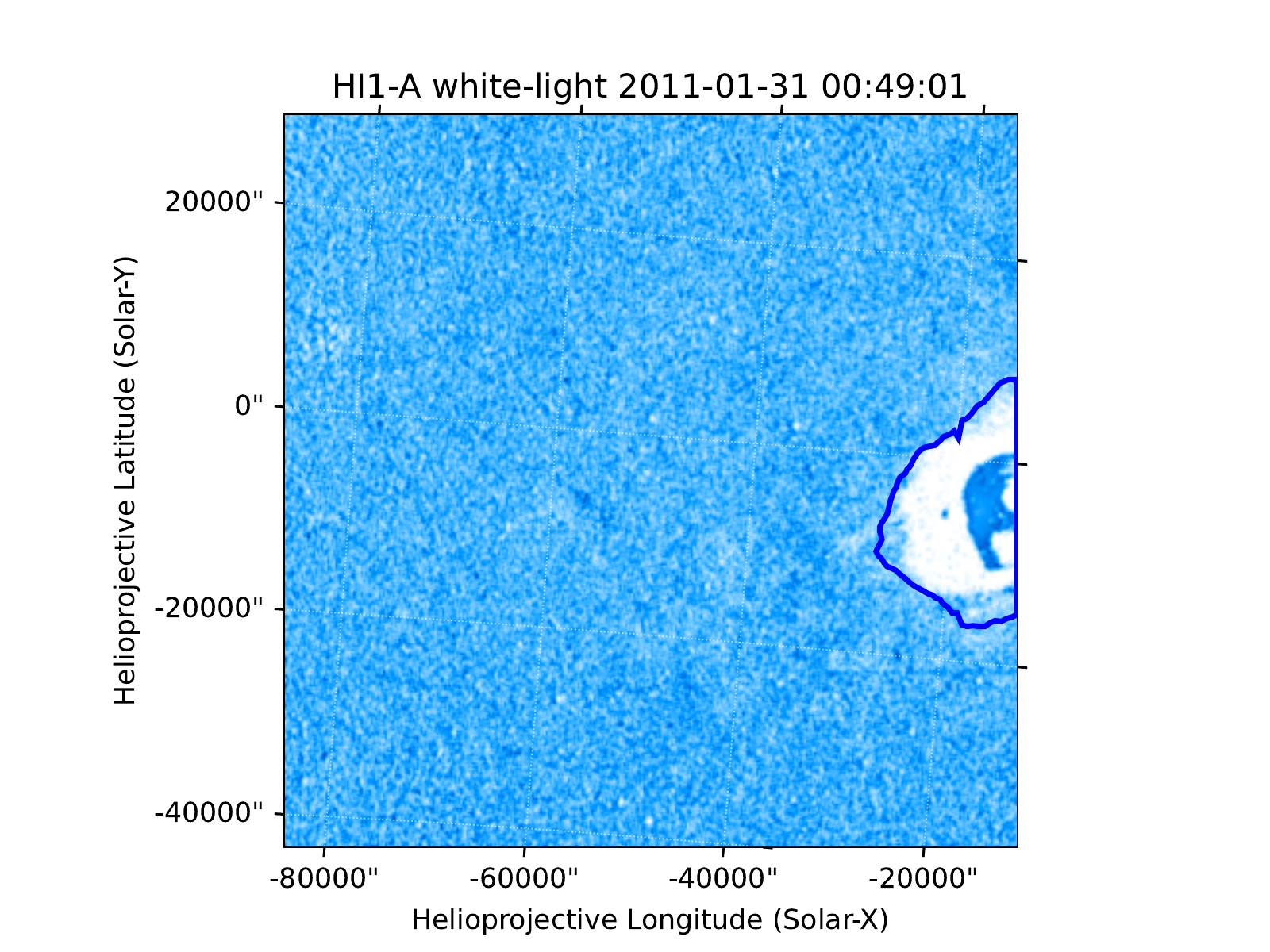}
  \kern-0.8cm
  \includegraphics[scale=0.2]{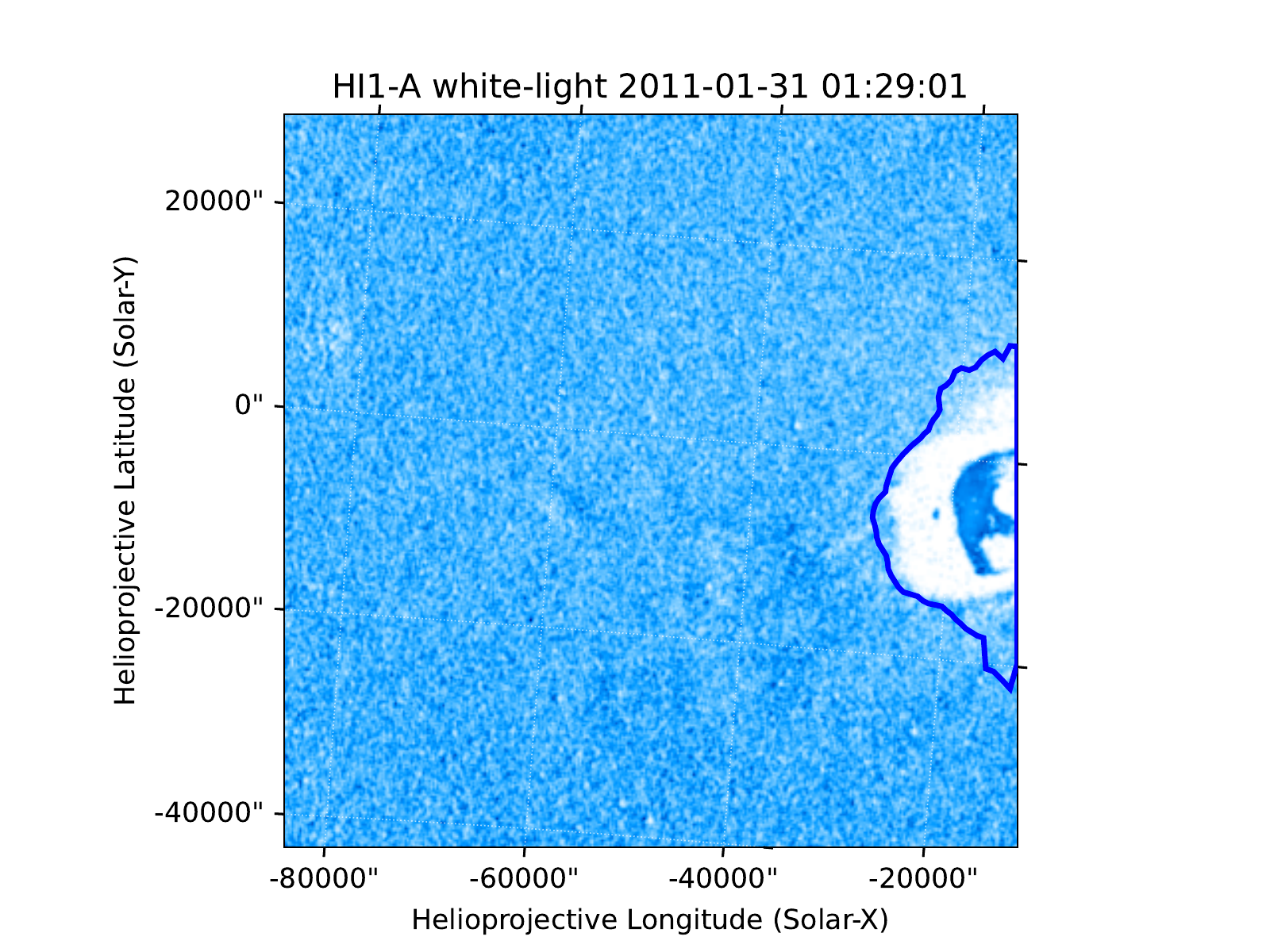}
  \kern-0.8cm
  \includegraphics[scale=0.2]{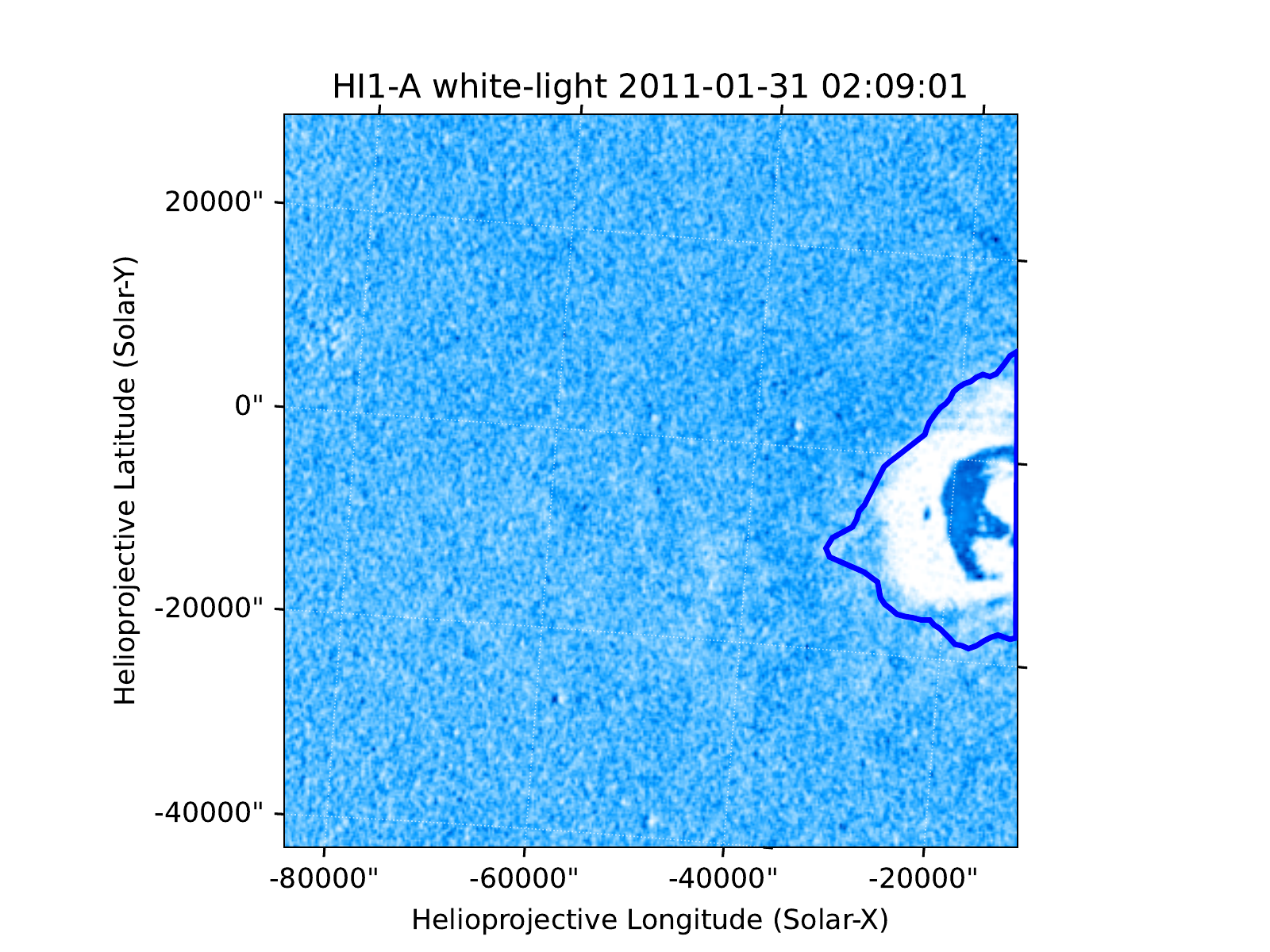}
}
\hbox{
  \includegraphics[scale=0.2]{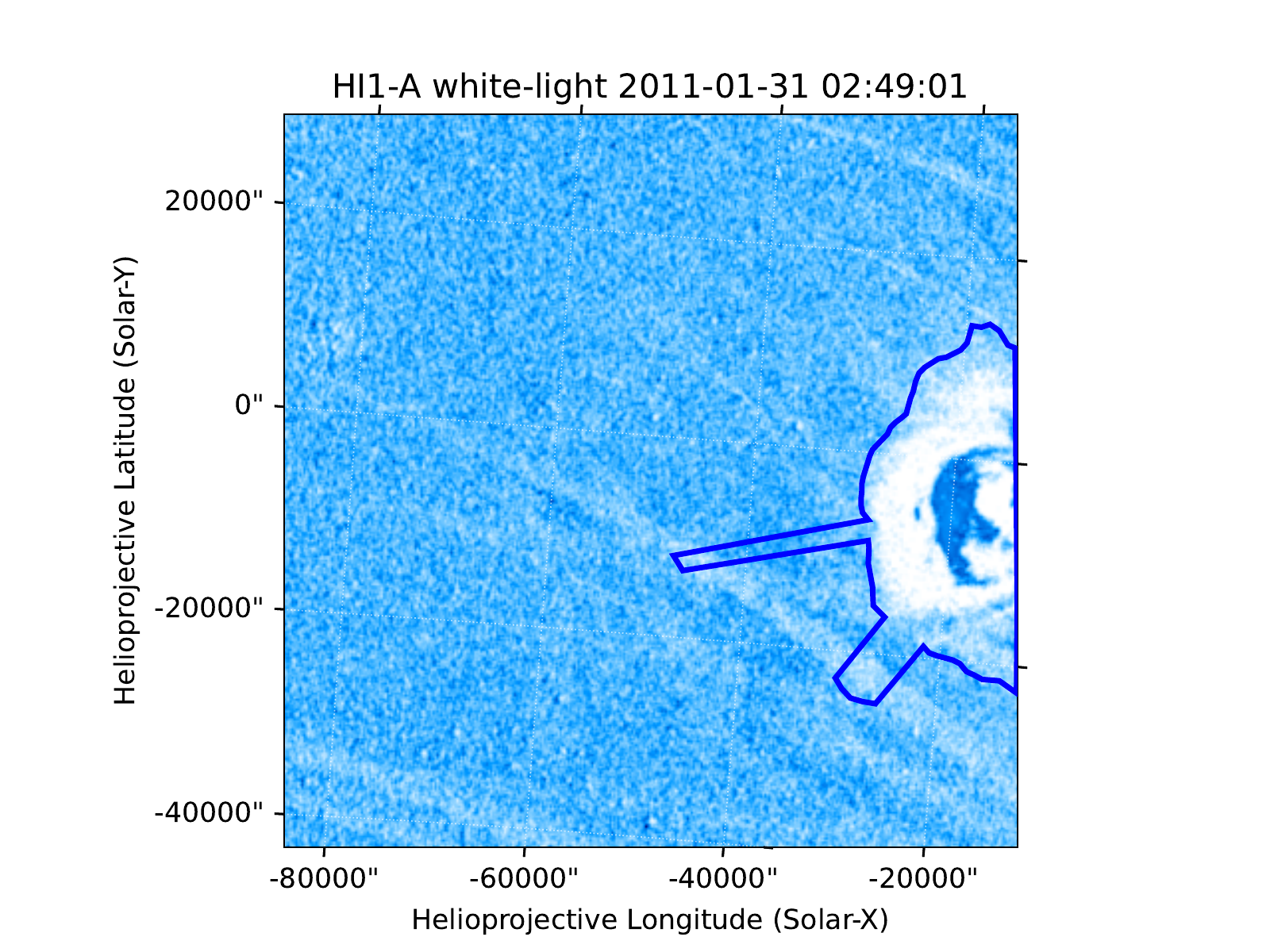}
  \kern-0.8cm
  \includegraphics[scale=0.2]{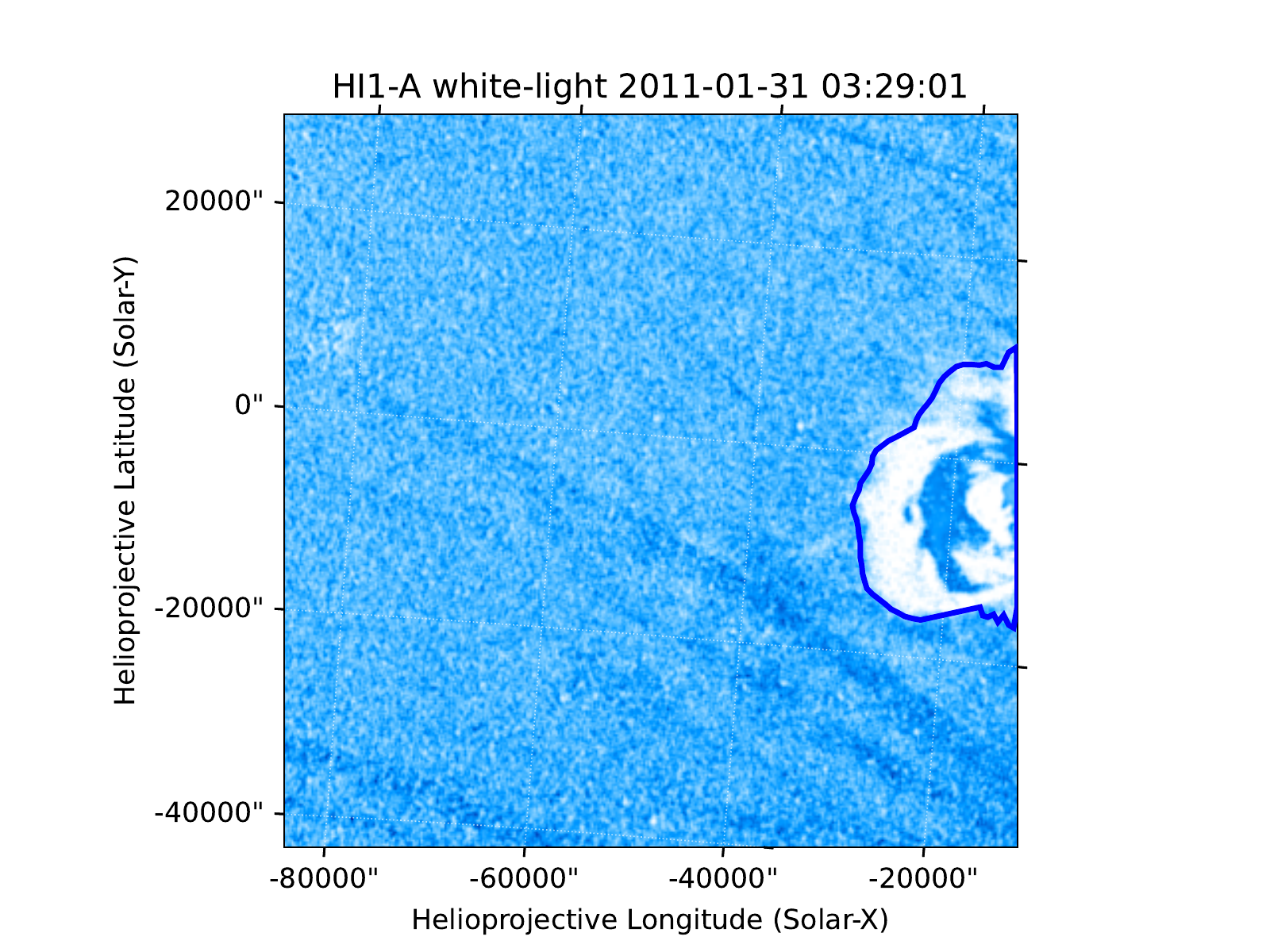}
  \kern-0.8cm
  \includegraphics[scale=0.2]{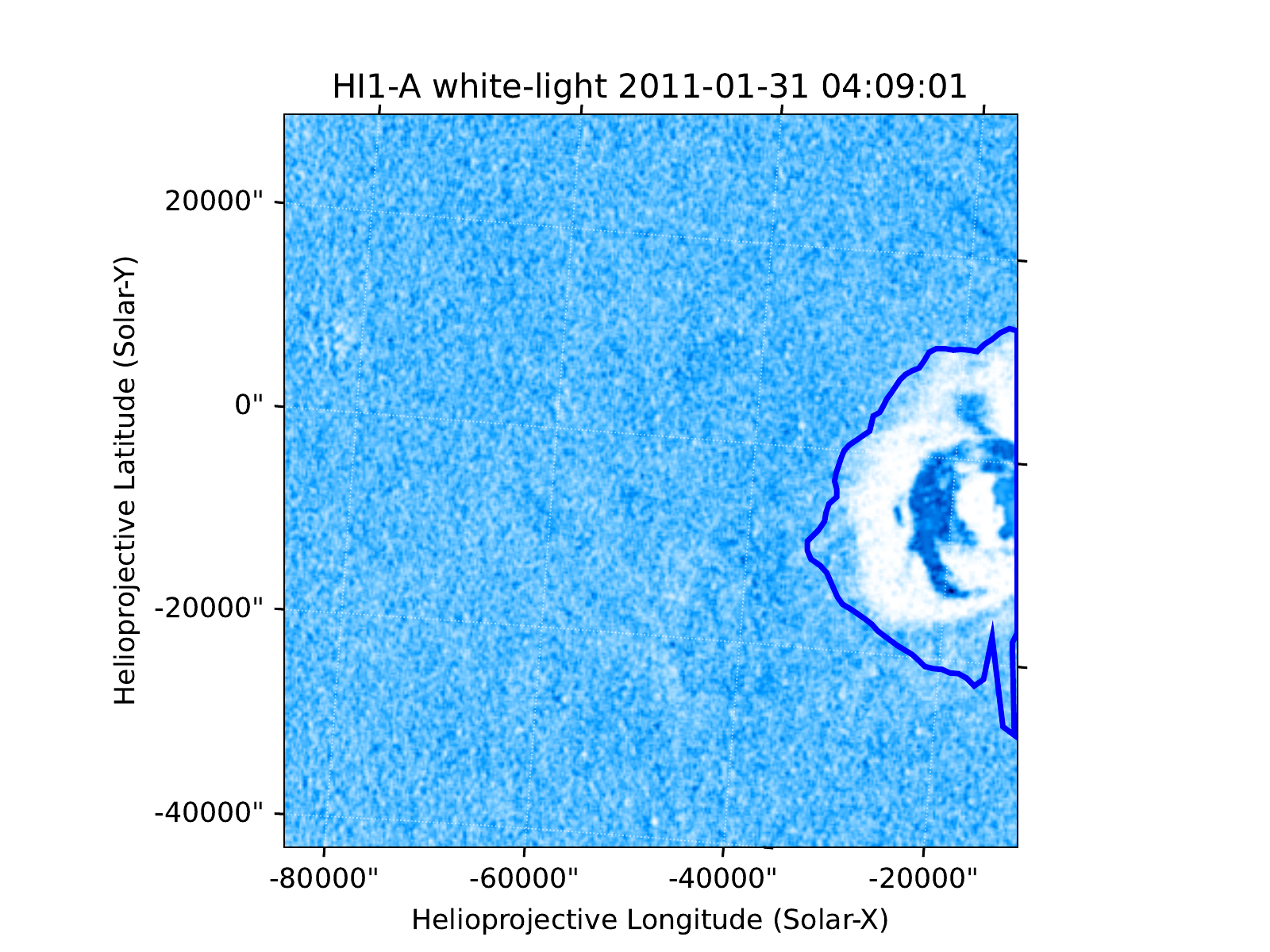}
}
  \caption{Longer sequence: successive images of a CME event advancing in the field of vision of the STEREO-A camera. The front speed can be estimated from such consecutive pairs and then averaged throughout, as explained in the Results section.}\label{Fig:Sequence}
\end{center}
\end{figure}

We have tracked CME events using a set of 100 STEREO 11-day background-removed L2 HI-1 a and b images from the UK Solar System Data Center, \url{https://www.ukssdc.ac.uk}. The algorithm performed with overall high tracking capabilities and without any fine-tuning or training set. We also conducted a manual identification of the CME regions, and then compared it with the automatic one.
We counted the relative differences in areas between the two perimeters as errors, and we measured by hand that the percentage
error of region identifications was below 15\% for the most irregular events. We did
not find false positives. The images in Figure~\ref{Fig:ProcExamples} and Figure~\ref{Fig:Sequence}
illustrate these cases.

\begin{figure}
\begin{center}
\hbox{\kern-0.5cm 
  \includegraphics[scale=0.35]{batch_b_20081212_180901}
  \kern-1.2cm
  \includegraphics[scale=0.35]{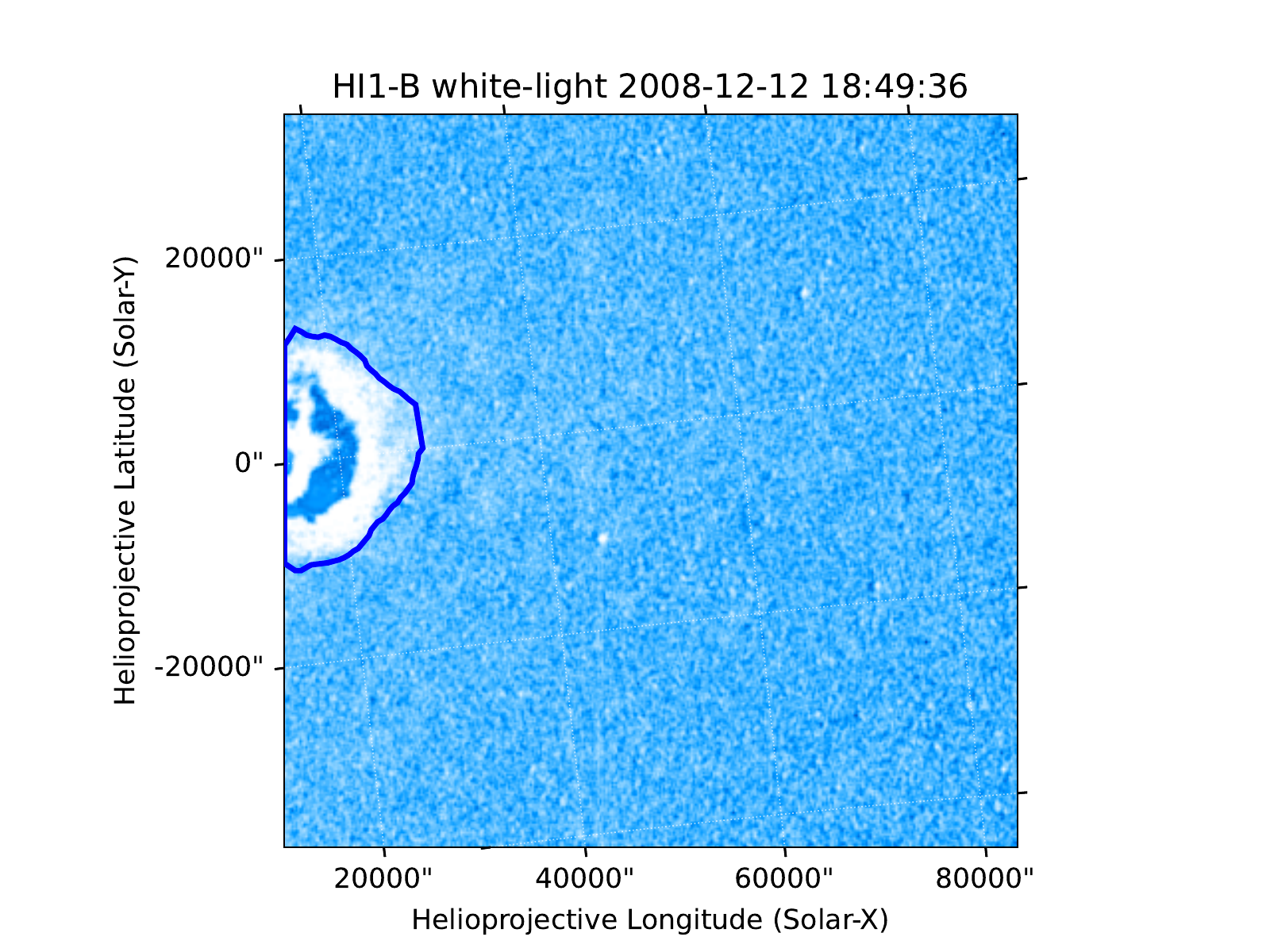}
}
\end{center}
\caption{Two-image sequence to estimate front speed, defined as the speed at which the outermost edge advances from a snapshot to another one. The calculation is detailed in the Results section.}\label{Fig:ShortSequence}
\end{figure}

A simplified
initial calculation appears in agreement with speeds of the order of $300$km/s from
\cite{Davis2009GeoRL..36.8102D} for the 2008 December 12 CME event, as effectively measured
by the SWEPAM instrument aboard the Advanced Composition Explorer spacecraft \cite{StoneACE1998SSRv...86....1S}. We used the
transformation (4) described in \cite{Thompson2006CoordSolIm}:
\[
  x\simeq D_{\odot}\left(\frac{\pi}{180^{\circ}}\right)\theta_x
  \]
to compute the displacement of the front, from the Helioprojective longitude displacement of 
$\theta_x = 0.27^{\circ}=1000$ seconds of arc from 18:09 to 18:49 of two algorithm-generated contours for HI-1 B, shown in Figure~\ref{Fig:ShortSequence}.
Thus, $\Delta x=730000$km, and $\Delta t=2400s$, so $v=300$km/s, using only two significant digits.

Succesive images, such as in Figure~\ref{Fig:Sequence} can allow more accurate estimations of kinematics. 

\section{Discussion}\label{Section:Discussion}

To improve the smoothness and completeness of the perimeter detection, we have also tried
a subsequent stage of the morphological geodesic active
contour method from \cite{MarquezNeilaMorphActiveContours2014}, based on level sets seeded with the interior
of an initially-found contour. So far, it did not seem to succeed, mainly because of the
of the various bright-dark alternating inner fronts, and also because of the background noise
from the light of the stars. The latter could be further reduced by carefully
realigning the images before subtraction, as suggested by \cite{Davies2009GeoRL..36.2102D}.

A possible line of improvement would be to better adapt the active contour methods to the already-found contour to
fine-align it along less-contrasting borders. A more precise way to distinguish the far-away background
could also help.
Finally, I am planning on working to integrate it with existing models to estimate CME velocities
and 3D shape changes, and then systematically comparing it with other methods. 

\section{Acknowledgements}
We have implemented the algorithm in Python, and made it available at {\small\url{https://github.com/MDNich/CME-Image-Edge-Detection-in-Python-from-STEREO-data}}, being thus grateful to the community maintaining Numpy, SciPy and Matplotlib, as well as \cite{sunpy_community2020}. We thank the reviewers for their careful reading and helpful comments and for suggestions for applications of these techniques to missions such as PUNCH and Solar Orbiter.

\bibliographystyle{aa}
\bibliography{bib4simpleCMEEdgeImg}

\begin{thebibliography}{25}
\expandafter\ifx\csname natexlab\endcsname\relax\def\natexlab#1{#1}\fi

\bibitem[{{Byrne} {et~al.}(2009){Byrne}, {Gallagher}, {McAteer}, \&
  {Young}}]{Byrne2009A&A...495..325B}
{Byrne}, J.~P., {Gallagher}, P.~T., {McAteer}, R.~T.~J., \& {Young}, C.~A.
  2009, \aap, 495, 325

\bibitem[{Canny(1986)}]{Canny1986}
Canny, J. 1986, IEEE Transactions on Pattern Analysis and Machine Intelligence,
  8, 679

\bibitem[{Caselles {et~al.}(1993)Caselles, Catt\'e, Coll, \&
  Dibos}]{Caselles1993}
Caselles, V., Catt\'e, F., Coll, T., \& Dibos, F. 1993, Numer. Math., 66, 1

\bibitem[{Chan \& Vese(2001)}]{ChanVese2001}
Chan, T.~F. \& Vese, L.~A. 2001, IEEE Transactions on Image Processing, 10, 266

\bibitem[{Crifo(1983)}]{Crifo1983LoopOrBubble}
Crifo, F. 1983, Solar Physics, 83, 143

\bibitem[{{Davies} {et~al.}(2009){Davies}, {Harrison}, {Rouillard}, {Sheeley},
  {Perry}, {Bewsher}, {Davis}, {Eyles}, {Crothers}, \&
  {Brown}}]{Davies2009GeoRL..36.2102D}
{Davies}, J.~A., {Harrison}, R.~A., {Rouillard}, A.~P., {et~al.} 2009, \grl,
  36, L02102

\bibitem[{{Davis} {et~al.}(2009){Davis}, {Davies}, {Lockwood}, {Rouillard},
  {Eyles}, \& {Harrison}}]{Davis2009GeoRL..36.8102D}
{Davis}, C.~J., {Davies}, J.~A., {Lockwood}, M., {et~al.} 2009, \grl, 36,
  L08102

\bibitem[{Eyles {et~al.}(2007)Eyles, Davis, Harrison, Waltham, Halain, Mazy,
  Defise, Howard, Moses, Newmark, \& Plunkett}]{Eyles10.1117/12.732822}
Eyles, C., Davis, C., Harrison, R., {et~al.} 2007, in Solar Physics and Space
  Weather Instrumentation II, ed. S.~Fineschi \& R.~A. Viereck, Vol. 6689,
  International Society for Optics and Photonics (SPIE), 40 -- 52

\bibitem[{{Eyles} {et~al.}(2009){Eyles}, {Harrison}, {Davis}, {Waltham},
  {Shaughnessy}, {Mapson-Menard}, {Bewsher}, {Crothers}, {Davies}, {Simnett},
  {Howard}, {Moses}, {Newmark}, {Socker}, {Halain}, {Defise}, {Mazy}, \&
  {Rochus}}]{Eyles2009SoPh..254..387E}
{Eyles}, C.~J., {Harrison}, R.~A., {Davis}, C.~J., {et~al.} 2009, \solphys,
  254, 387

\bibitem[{Fisher(1984)}]{Fisher1984CMEEvents}
Fisher, R. 1984, Adv. Space Res., 4, 163

\bibitem[{Fisher \& Munro(1984)}]{Fisher1984CMEGeom}
Fisher, R. \& Munro, H.~R. 1984, Astrophys. J., 280, 163

\bibitem[{Gopalswamy \& Yashiro(2007)}]{Gopalswamy2007}
Gopalswamy, N. \& Yashiro, S. 2007, Journal of Geophysical Research, 112

\bibitem[{Howard {et~al.}(2008)Howard, Nandy, \& Koepke}]{Howard2008}
Howard, T., Nandy, D., \& Koepke, A. 2008, Journal of Geophysical Research, 113

\bibitem[{{Michalek} {et~al.}(2006){Michalek}, {Gopalswamy}, {Lara}, \&
  {Yashiro}}]{MichalekGopalswamy2006SpWea...410003M}
{Michalek}, G., {Gopalswamy}, N., {Lara}, A., \& {Yashiro}, S. 2006, Space
  Weather, 4, S10003

\bibitem[{Márquez-Neila {et~al.}(2014)Márquez-Neila, Baumela, \&
  Alvarez}]{MarquezNeilaMorphActiveContours2014}
Márquez-Neila, P., Baumela, L., \& Alvarez, L. 2014, IEEE Transactions on
  Pattern Analysis and Machine Intelligence, 36, 2

\bibitem[{{Olmedo} {et~al.}(2008){Olmedo}, {Zhang}, {Wechsler}, {Poland}, \&
  {Borne}}]{Olmedo2008AGUSMSP43A..02O}
{Olmedo}, O., {Zhang}, J., {Wechsler}, H., {Poland}, A., \& {Borne}, K. 2008,
  in AGU Spring Meeting Abstracts, Vol. 2008, SP43A--02

\bibitem[{{Robbrecht} \& {Berghmans}(2004)}]{Robbrecht2004A&A...425.1097R}
{Robbrecht}, E. \& {Berghmans}, D. 2004, \aap, 425, 1097

\bibitem[{{Sheeley} {et~al.}(1997){Sheeley}, {Wang}, {Hawley}, {Brueckner},
  {Dere}, {Howard}, {Koomen}, {Korendyke}, {Michels}, {Paswaters}, {Socker},
  {St. Cyr}, {Wang}, {Lamy}, {Llebaria}, {Schwenn}, {Simnett}, {Plunkett}, \&
  {Biesecker}}]{Sheeley1997ApJ...484..472S}
{Sheeley}, N.~R., {Wang}, Y.~M., {Hawley}, S.~H., {et~al.} 1997, \apj, 484, 472

\bibitem[{{Stone} {et~al.}(1998){Stone}, {Frandsen}, {Mewaldt}, {Christian},
  {Margolies}, {Ormes}, \& {Snow}}]{StoneACE1998SSRv...86....1S}
{Stone}, E.~C., {Frandsen}, A.~M., {Mewaldt}, R.~A., {et~al.} 1998, \ssr, 86, 1

\bibitem[{{The SunPy Community} {et~al.}(2020){The SunPy Community}, Barnes,
  Bobra, Christe, Freij, Hayes, Ireland, Mumford, Perez-Suarez, Ryan, Shih,
  Chanda, Glogowski, Hewett, Hughitt, Hill, Hiware, Inglis, Kirk, Konge, Mason,
  Maloney, Murray, Panda, Park, Pereira, Reardon, Savage, Sipőcz, Stansby,
  Jain, Taylor, Yadav, Rajul, \& Dang}]{sunpy_community2020}
{The SunPy Community}, Barnes, W.~T., Bobra, M.~G., {et~al.} 2020, The
  Astrophysical Journal, 890, 68

\bibitem[{{Thompson}(2006)}]{Thompson2006CoordSolIm}
{Thompson}, W.~T. 2006, Astronomy and Astrophysics, 449, 791

\bibitem[{{Webb} \& {Howard}(2012)}]{WebbHowardCMEObsReview2012}
{Webb}, D.~F. \& {Howard}, T.~A. 2012, Living Reviews in Solar Physics, 9

\bibitem[{{Xie} {et~al.}(2006){Xie}, {Gopalswamy}, {Ofman}, {St. Cyr},
  {Michalek}, {Lara}, \& {Yashiro}}]{XieGopalswamy2006SpWea...410002X}
{Xie}, H., {Gopalswamy}, N., {Ofman}, L., {et~al.} 2006, Space Weather, 4,
  S10002

\bibitem[{{Yashiro} {et~al.}(2004){Yashiro}, {Gopalswamy}, {Michalek}, {St.
  Cyr}, {Plunkett}, {Rich}, \& {Howard}}]{2004JGRA..109.7105Y}
{Yashiro}, S., {Gopalswamy}, N., {Michalek}, G., {et~al.} 2004, Journal of
  Geophysical Research (Space Physics), 109, A07105

\bibitem[{{Young} \& {Gallagher}(2008)}]{Young2008SoPh..248..457Y}
{Young}, C.~A. \& {Gallagher}, P.~T. 2008, \solphys, 248, 457

\end{thebibliography}

\end{document}